\documentclass[9pt,twocolumn,twoside]{osajnl}

\journal{arxiv} 

\setboolean{shortarticle}{true}

\title{Dual-microcomb generation in a synchronously-driven waveguide ring resonator}

\author[1,2,4]{Yiqing Xu}
\author[1,2]{Miro Erkintalo}
\author[3]{Yi Lin}
\author[1,2]{St\'ephane Coen}
\author[3,5]{Huilian Ma}
\author[1,2]{Stuart G. Murdoch}

\affil[1]{The Dodd-Walls Centre for Photonic and Quantum Technologies, Auckland 1010, New Zealand}
\affil[2]{Department of Physics, University of Auckland, Auckland 1010, New Zealand}
\affil[3]{School of Aeronautics and Astronautics, Zhejiang University, 310027, Hangzhou, China}
\affil[4]{e-mail: yxu079@aucklanduni.ac.nz}
\affil[5]{e-mail: mahl@zju.edu.cn}

\begin{abstract}
Microcombs -- optical frequency combs generated in coherently-driven nonlinear microresonators -- have attracted significant attention over the last decade. The ability to generate two such combs in a single resonator device has in particular enabled a host of applications from spectroscopy to imaging. Concurrently, novel comb generation techniques such as synchronous pulsed driving have been developed to enhance the efficiency and flexibility of microcomb generation. Here we report on the first experimental demonstration of dual-microcomb generation via synchronous pulsed pumping of a single microresonator. Specifically, we use two electro-optically generated pulse trains derived from a common continuous wave laser to simultaneously drive two orthogonal polarization modes of an integrated silica ring resonator, observing the generation of coherent dissipative Kerr cavity soliton combs on both polarization axes. Thanks to the resonator birefringence, the two soliton combs are associated with different repetition rates, thus realizing a dual-microcomb source. To illustrate the source's application potential, we demonstrate proof-of-concept spectroscopic measurements.
\end{abstract}

\setboolean{displaycopyright}{true}

\begin{document}

\maketitle

\noindent The generation of coherent optical frequency combs (``microcombs'') in nonlinear microresonators has attracted significant attention over the last decade~\cite{DelHaye2007,DelHaye2011,Kippenberg2011,Pasquazi2018,Gaeta2019}, enabling advances across a number of fields from astronomy to telecommunications~\cite{Marin-Palomo2017,Obrzud2019,Suh2019}. The fact that a single microresonator device can be used for the simultaneous generation of two (or even three) microcombs with different line spacing has in particular allowed for the realization of novel sources of dual-combs, whose many applications include ultra-high-speed optical ranging, spectroscopy, and imaging~\cite{Bao2019,Suh2016,Suh2018,Trocha2018,Dutt2018}.

In the most common paradigm of microcomb generation, a single microcresonator is coherently-driven with a single continuous wave (cw) laser~\cite{DelHaye2007,DelHaye2011,Kippenberg2011}. By judiciously controlling the detuning between the driving laser and a cavity resonance, it is possible to realize coherent microcombs that correspond to so-called dissipative Kerr cavity solitons -- pulses of light that can persist in the resonator without distortion~\cite{Leo2010,Coen2013,Coen2013_2,Herr2013,Kippenberg2018}. Whilst appealing in its simplicity, the use of a cw laser suffers from several deficiencies, including low pump-to-comb power conversion efficiency and difficulty to control the number and relative positions of excited solitons~\cite{Cole2017,Cole2018,Karpov2019,Guo2017}. These deficiencies generically extend to the domain of dual-microcomb generation, which typically relies on the use of two (counter-propagating, frequency-shifted, or orthogonally polarised) cw signals to drive two different resonator mode families associated with different free-spectral ranges (FSRs)~\cite{Suh2016,Suh2018,Trocha2018,Bao2019}.

More recent works have shown that coherent soliton microcombs can also be generated by driving a resonator with a train of pulses whose repetition rate is close to the resonator FSR (or a rational fraction of the FSR)~\cite{Obrzud2017,Xu2020}. Compared to cw pumping, such a synchronous driving scheme has potential to enhance pump-to-comb conversion efficiency, and it permits an improved capacity for the controllable generation of single-soliton microcombs. Moreover, because the generated microcomb is intrinsically synchronized to the repetition rate of the driving pulse train, which can be readily derived from (and hence controlled by) an external clock signal, synchronous driving offers an inherent time-frequency reference for the generated comb. Somewhat surprisingly, whilst these advantages would be of particular significance for dual-comb applications, demonstrations of synchronously-driven dual-microcombs have remained elusive.

In this Letter, we report on the experimental demonstration of a synchronously-driven dual-microcomb source that is based on soliton microcomb generation in a single optical waveguide ring resonator (WRR). Using two electro-optically generated pulse trains derived from a single narrow-linewidth cw laser, we simultaneously excite soliton combs in the two orthogonal polarization mode families of the resonator. Thanks to the birefringence of the WRR, the two polarization mode families exhibit different FSRs, thus allowing for the simultaneous generation of two soliton combs with different line spacings that are synchronized to the repetition rates of the respective driving pulse trains. The realized dual-comb source spans more than 20~nm around 1550~nm with a mean comb line spacing of 3.23~GHz and a spacing difference of 84~kHz. We demonstrate the applied potential of the source by performing proof-of-concept spectroscopic measurements, though note that such a source is also ideally suited for other dual-comb applications including ranging and imaging.

\begin{figure}[!htbp]
\centering
  \includegraphics[width = \linewidth, clip = true]{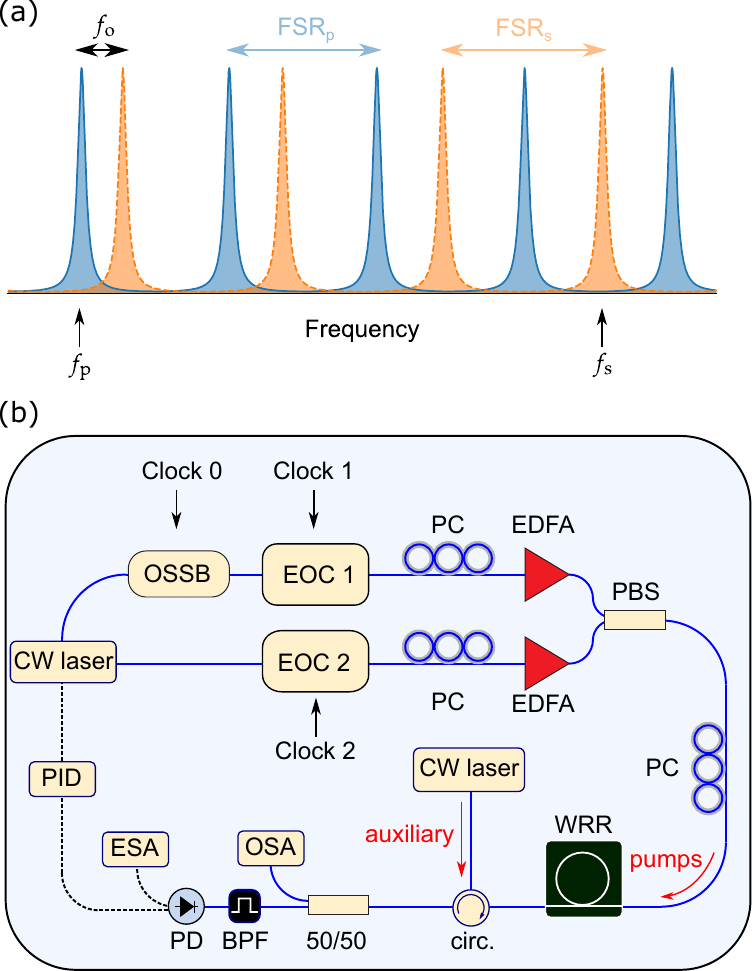}
  \caption{(a) Schematic illustration of the mode spectrum of our resonator. Resonances of the primary mode family (blue) are spaced by $\text{FSR}_\mathrm{p}$ whilst those of the orthogonally-polarised secondary mode family (orange) are spaced by $\text{FSR}_\mathrm{s}$. Around the primary pump frequency $f_\mathrm{p}$ the resonances are offset by $f_\mathrm{o}\approx 426~\mathrm{MHz}$. The secondary-mode family is pumped at \mbox{$\Delta f=f_\mathrm{o} + 3\text{FSR}_\mathrm{s}$}. (b) Schematic illustration of the experimental setup. PC, polarization controller; circ., circulator; PID, proportional-integral-derivative; BPF, band-pass filter; PD, photodetector.}
  \label{fig1}
\end{figure}

We first describe the mode structure of our resonator so as to elucidate the mechanism of our driving scheme [see Fig.~\ref{fig1}(a)]. The waveguide that forms the resonator supports a single spatial mode such that the WRR only exhibits two mode families that correspond to two orthogonal polarizations (referred below as the primary and secondary mode families). Because of residual birefringence, the two mode families exhibit different FSRs with a mean of $\overline{\text{FSR}} = (\text{FSR}_\mathrm{p} + \text{FSR}_\mathrm{s})/2\approx 3.23~\mathrm{GHz}$ and difference $\Delta\text{FSR}=\text{FSR}_\mathrm{s} - \text{FSR}_\mathrm{p}\approx84~\mathrm{kHz}$, where $\text{FSR}_\mathrm{p} = \overline{\text{FSR}}-\Delta\text{FSR}/2$ and $\text{FSR}_\mathrm{s} = \overline{\text{FSR}}+\Delta\text{FSR}/2$ are the FSRs of the primary and secondary mode families, respectively.  Moreover, the resonances of the two mode families are offset by about $f_\mathrm{o} = 426~\mathrm{MHz}$ at the pump wavelength of 1550~nm.

Figure~\ref{fig1}(b) shows our experimental setup. The WRR is fabricated from a Ge-doped low-loss silica-on-silicon waveguide~\cite{Zhang2017,Lin2019,Xu2020}. It is designed to support a single spatial mode in the telecommunication band and has a measured finesse of 340, corresponding to 9.5~MHz resonance width. The WRR is driven by two orthogonally polarised pulse trains that are carved from a single narrow-linewidth cw laser at 1550~nm (Koheras Adjustik E15, linewidth $< 1~\mathrm{kHz}$). To allow for a single cw laser to be used as a common source for both of modes, we split the cw laser light into two paths and use optical single-sideband~(OSSB) modulation to down-shift the frequency of light along the second paths by $\Delta f = f_\mathrm{o} + 3\text{FSR}_\mathrm{s} =10.1~\mathrm{GHz}$~\cite{Hraimel2011}. (Note: the frequency shift includes a three-fold multiple of the $\text{FSR}_\mathrm{s}$ simply because a larger frequency shift is easier to achieve with the OSSB modulator we had at our disposal.)

To convert the cw beams into pulse trains, we use two independent electro-optic comb (EOC) generators [labelled EOC~1 and EOC~2 in Fig.~\ref{fig1}(b)]. Both EOC generators use one amplitude and two phase modulators that are driven by independent RF clocks [Clock~1 and Clock~2 in Fig.~\ref{fig1}(b)], followed by cascaded linear and nonlinear compression. The resulting pulse trains were measured to have a full-width half maximum~(FWHM) of 1.8~ps (EOC~1) and 2.5~ps (EOC~2), with the difference resulting from the different modulators used in the comb generators. By precisely adjusting the RF clock signals applied on the EO comb generators to match the FSRs of the orthogonally polarised cavity modes, we achieve independent, synchronous pumping of both modes. After their generation, and prior to being launched into the WRR, the two EO combs are amplified with two erbium-doped fiber amplifiers (EDFAs) and combined with a polarising beam splitter (PBS). The average power of EOC~1 and~2 was measured to be 10~mW and~20~mW, respectively, at the input of the WRR. A polarization controller is used just before the WRR to align the orthogonal pulse trains along each of the orthogonal WRR cavity modes.

To facilitate soliton comb generation, we use an auxiliary cw laser at 1530~nm that counter-propagates with respect to the main pumps to provide thermal compensation and stabilization~\cite{Carmon2014}. With the auxiliary laser thermally self-stabilized to a cavity resonance, soliton combs can be generated simply by tuning the main 1550~nm cw pump laser into resonance. By controlling the RF clock applied to the OSSB stage [Clock~0 in Fig.~\ref{fig1}(b)], we ensure soliton combs are simultaneously generated along each of the orthogonal cavity modes. We detect the total intensity at the cavity output using a slow photodetector and lock the resulting signal at a set level using a PID controller that actuates the wavelength of the 1550~nm pump laser. Once the soliton combs are generated and stabilized in this manner, we characterize them using an optical spectrum analyzer (OSA) and an electronic spectrum analyzer (ESA) with a 5~GHz photodetector.

\begin{figure}[!t]
\centering
  \includegraphics[width = \linewidth, clip = true]{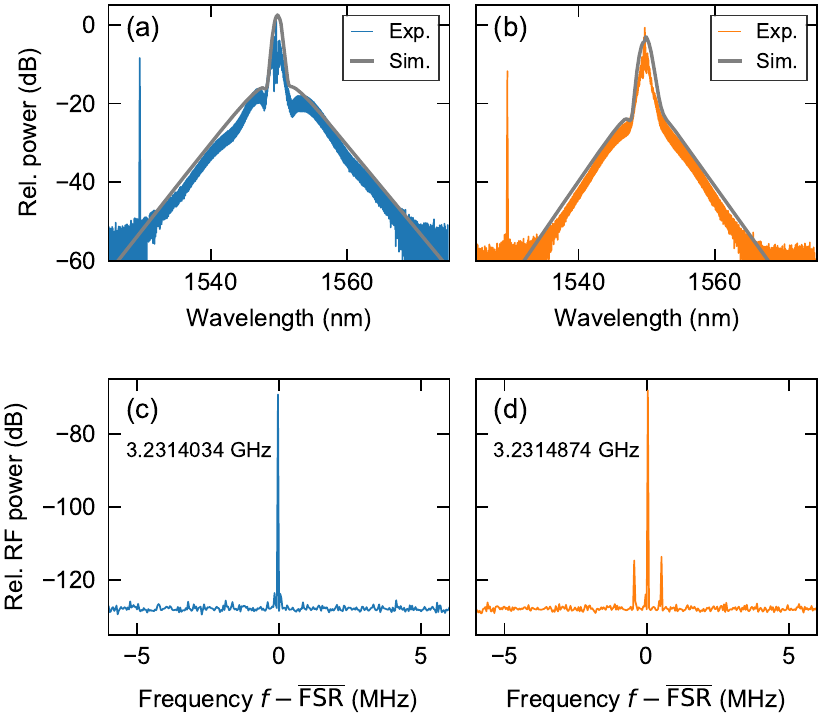}
  \caption{(a, b) Measured (blue and orange) and simulated (gray) optical spectra of cavity soliton combs generated in isolation via synchronous pumping of the (a) primary and (b) secondary polarization modes of the WRR. The simulations use experimental parameters with cavity phase detunings estimated by fitting to experimental spectra: (a) $\delta_0 = 0.33~\mathrm{rad}$; (b) $\delta_0 = 0.21~\mathrm{rad}$. (c) and (d) show the corresponding RF spectra around the fundamental beat tones. Resolution bandwidth in (c) and (d) was 10~Hz.}
  \label{fig2}
\end{figure}

Figure~\ref{fig2}(a) and (b) show the optical spectra of the soliton combs when the two combs are generated independently (with only one electro-optically generated pulse train pumping the cavity), whilst Fig.~\ref{fig2}(c) and (d) show corresponding RF spectra around the fundamental harmonic centred at $\overline{\text{FSR}}$. The optical bandwidth of the secondary comb [Fig.~\ref{fig2}(b)] is slightly smaller than that of the primary comb [Fig.~\ref{fig2}(a)], which we attribute to slightly different cavity detunings associated with the two different modes. (This hypothesis is corroborated by numerical simulations of the Lugiato-Lefever model that use experimental parameters but with the two detunings considered free parameters~\cite{Coen2013_2}.) The cw component visible around 1530~nm in the optical spectra arises from the back-scattering of the auxiliary cw laser used for thermal stabilization. The RF beat tones of the combs are precisely identical to the RF clock frequencies used to generate the pump pulse trains, and offset from one another by $\Delta \text{FSR} \approx 84~\mathrm{kHz}$. Moreover, as expected for soliton combs, the RF spectra show low noise, indicating coherent comb states. We note, however, that for the secondary comb, low-amplitude sidebands can clearly be observed in the RF spectrum. These are not indicative of any soliton breathing instabilities, but arise rather due to imperfections in electronic amplifiers that drive the modulators used to create the corresponding pump pulse train. Indeed, we find that these sidebands are present even before that pump pulse train is launched into the WRR. This artifact does not impact on the dual-comb source to be reported next.

\begin{figure}[!t]
\centering
  \includegraphics[width = \linewidth, clip = true]{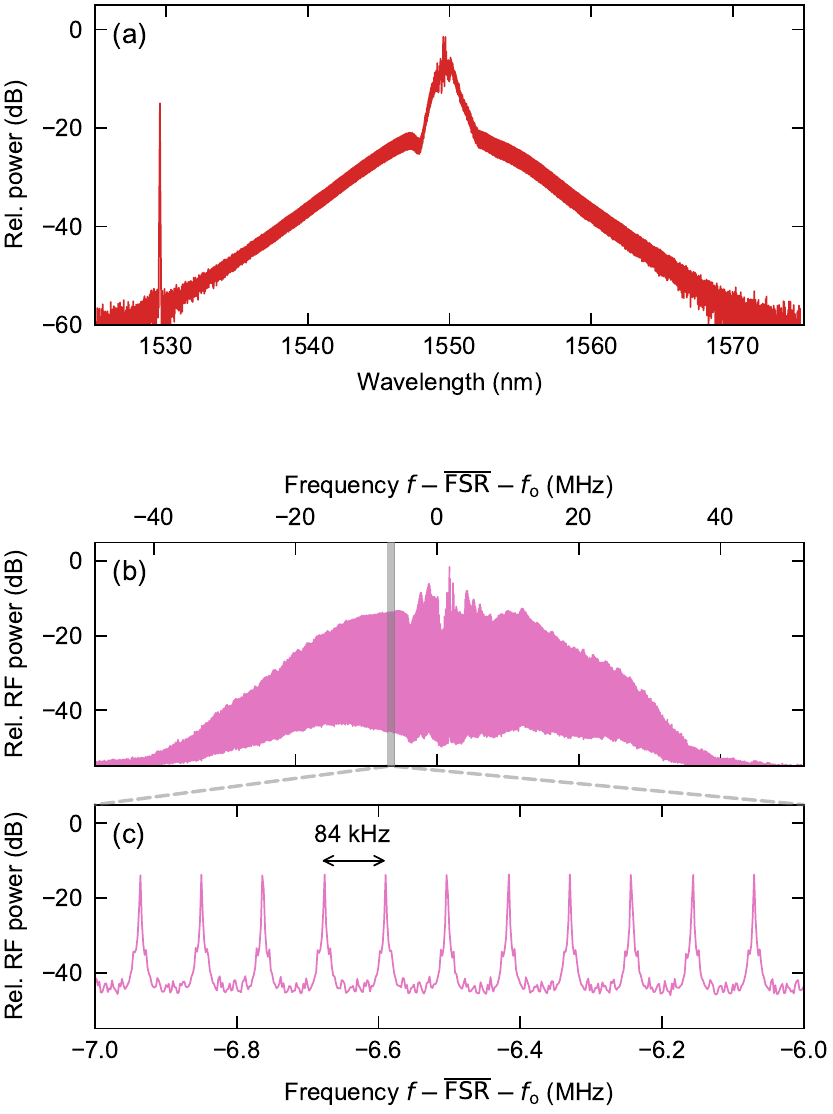}
  \caption{(a) Optical spectrum of the dual-microcomb when solitons combs are simultaneously generated along both polarization modes via synchronous pumping. (b) RF spectrum measured around $f_\mathrm{o}+\overline{\text{FSR}}$ after the dual-microcomb was passed through a waveshaper to suppress the pump components. (c) A zoomed inset around a 1-MHz-wide segment of the dual-comb RF spectrum, revealing comb lines separated by $\Delta\text{FSR}=84~\mathrm{kHz}$.}
  \label{fig3}
\end{figure}

By launching both pump pulse trains into the WRR, we are able to generate the two soliton combs simultaneously. The total optical spectrum of the resulting state, shown in Fig.~\ref{fig3}(a), is similar to the spectra observed when only one of the pump pulse trains is launched into the cavity [c.f. Fig.~\ref{fig2}(a) and (b)]. To facilitate the characterization and spectroscopic measurements that will follow, we pass the dual-comb state through an optical waveshaper to suppress the strong residual pump components before the dual-comb is analyzed with the ESA.

Thanks to the comparatively small FSR of the WRR, our ESA has sufficient electronic bandwidth to resolve several orders of the RF combs centred at $f_o + n\overline{\text{FSR}}$. In Fig.~\ref{fig3}(b) and~(c), we show the measured RF comb around the fundamental beat tone at $f_o+\overline{\text{FSR}}=3.66~\mathrm{GHz}$. The RF spectrum contains more than 400 lines within 6-dB electronic power level, with over 800 RF lines observed in total. This corresponds to a measured optical bandwidth of over 20~nm. Moreover, as can be seen in Fig.~\ref{fig3}(c), the spacing between the adjacent RF comb lines is precisely the 84~kHz difference between Clocks 1 and 2 used to generate the respective pulse trains.

\begin{figure}[!t]
\centering
  \includegraphics[width = \linewidth, clip = true]{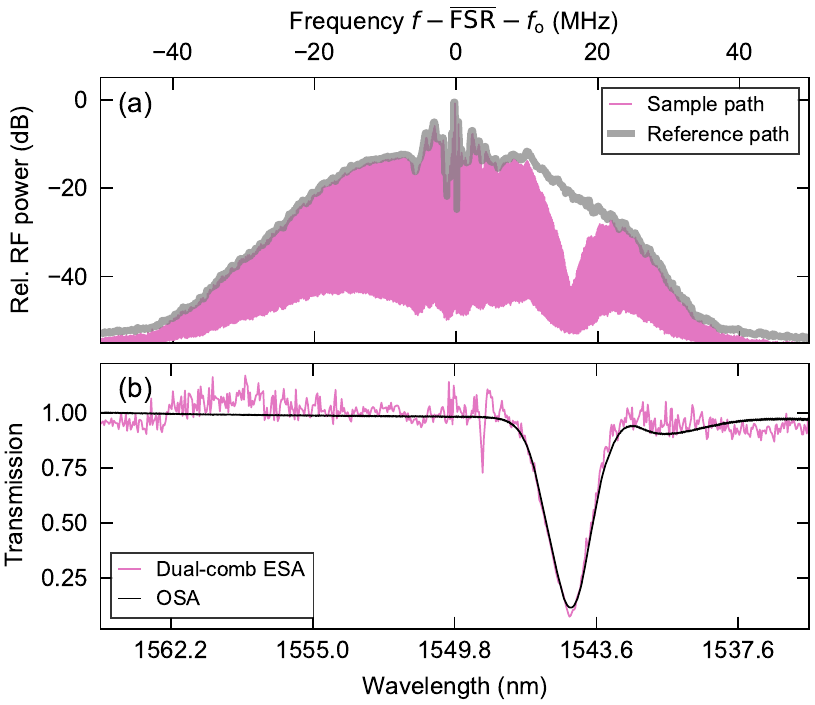}
  \caption{(a) Pink curve shows the RF spectrum of the dual-microcomb after the optical comb has passed through an FBG with 250~GHz reflection bandwidth, whilst the gray curve shows the corresponding reference RF spectrum. (b) Pink curve shows the normalized transmission spectrum for the FBG as extracted from the dual-comb data, whilst the black curve shows the corresponding result from direct optical transmission measurements performed with an OSA.}
  \label{fig4}
\end{figure}

To the best of our knowledge, the results reported in Fig.~\ref{fig3} correspond to the first dual-microcomb source achieved with pulsed pumping. To demonstrate the source's application potential, we use it to measure the spectral transmission profile of a fiber Bragg grating (FBG) with a center wavelength of 1545~nm and a reflection width of 250~GHz (FWHM). For this measurement, the dual-comb is first amplified with another EDFA and then split into two paths: one of the paths passes through the sample whilst the other is used as a reference. The pink curve in Fig.~\ref{fig4}(a) shows the RF spectrum after the dual comb has passed through the FBG, whilst the gray curve shows the corresponding reference RF spectrum. The impact of the FBG is evident on the RF spectrum of the sample path, with tens of beat tones attenuated. Figure~\ref{fig4}(b) shows the normalized transmission profile as extracted from the RF spectra (pink curve) compared with the optical transmission profile (black). As can be seen, the two curves are in very good agreement. The spectral resolution of this measurement is 26~pm which corresponds to the $\overline{\text{FSR}}$ of the WRR.

In summary, we have presented an experimental demonstration of a synchronously-driven dual-microcomb. By driving two orthogonal polarization modes of an integrated waveguide ring resonator with two electro-optically generated pulse trains with different repetition rates, we have been able to generate two coherent soliton combs with different line spacing. Moreover, we have demonstrated the source's application potential via proof-of-concept spectroscopic measurements. The use of synchronous pulsed pumping ensures operation in the single-soliton regime and provides a direct time/frequency reference for both generated combs thanks to their synchronization with the electronically generated pulse trains. Finally, whilst our proof-of-concept demonstrations have focussed on the use of synchronous pumping, our scheme can also be extended to rational harmonic driving so as to achieve optical combs with larger line spacing if desired~\cite{Xu2020}.

\section*{Funding}
Marsden Fund and the Rutherford Discovery Fellowships of the Royal Society of New Zealand. National Natural Science Foundation of China (62175218).

\section*{Disclosures}
\noindent The authors declare no conflicts of interest.\\

\end{document}